\documentstyle[prl,aps,twocolumn]{revtex}
\title{Microscopic theory of Josephson mesoscopic constrictions}
\author{A. Mart\'{\i}n-Rodero, F. J. Garc\'{\i}a-Vidal and A. Levy Yeyati}
\address{
Departamento de F\'\i sica de la Materia Condensada C-XII.\\
Facultad de Ciencias. Universidad Aut\'onoma de Madrid.\\
E-28049 Madrid. Spain.}
\begin{document}

\draft
\maketitle

\begin{abstract}
We present a microscopic theory for the d.c. Josephson effect in model
mesoscopic constrictions. Our method is based on a non-equilibrium Green
function formalism which allows for a self-consistent determination of
the order parameter profile along the constriction. The various regimes
defined by the different length scales (Fermi wavelength $\lambda_F$,
coherence length $\xi_0$ and constriction length $L_C$)
can be analyzed, including the case where all these lengths are
comparable. For the case $\lambda_F  \tilde{<} (L_C,\xi_0)$
phase oscillations with
spatial period $\lambda_F/2$ can be observed. In the case of $L_C>\xi_0$
solutions with a phase-slip center inside the constriction can be found,
in agreement with previous phenomenological theories.
\end{abstract}

\vspace{0.3in}

\narrowtext

The problem of d.c. Josephson transport through a superconducting
constriction received a considerable attention several years ago in the
context of what was known as the theory of superconducting weak-links
\cite{Likharev}.
In the last few years, there has been a renewed interest on the subject
fostered by the advances in nano-scale technologies which would allow
the fabrication of superconducting mesoscopic devices. Actually, some
steps in this direction have already been taken with the development
of a Josephson Field Effect Transistor (JOFET) \cite{Exp}. Another
experimental
situtation where this problem can be studied is that of a superconducting
Scanning Tunneling Microscope \cite{STM}.

The relevant feature of such a mesoscopic device would be the phase
coherence of both single-electrons and Cooper pairs over a length
comparable to the system size. This opens the possibility of observing
novel interference phenomena when $\lambda_F \leq(\xi_0,\xi_N,L_C)$ ($\xi_N$
denotes the normal electrons coherence length).
The case in which $L_C \ll \xi_0$ has been analyzed in a recent publication
by Beenakker and van Houten \cite{Beenakker}. In this regime the detailed
form of the order parameter profile inside the constriction is irrelevant
for the evaluation of the Josephson current. In the opposite limit ($L_C \gg
\xi_0$) the superconducting phase would drop linearly along the
constriction. A model for a superconducting point contact within this
limit has been proposed by Furusaki et al \cite{Furusaki}.

In the intermediate regime, in which all the relevant lengths can be
comparable, one would need a self-consistent determination of the complex
order parameter along the constriction. This poses a difficult task that
has only been addressed within phenomenological Ginzburg-Landau theory
\cite{Ginzburg} or with semi-classical treatments based on Boltzman-type
transport equations \cite{Kulik}. The aim of this work is to present
a microscopic model for the self-consistent description of a Josephson
constriction in all the relevant regimes.

We consider a model constriction like the one depicted in Figure 1, which
consists on a quasi one dimensional region of length $L_C$ coupled to wider
regions ($L$ and $R$) which act as electron reservoirs. For simplicity, we
assume a single quantum channel open for transport through the
constriction (a generalization for the multi-channel case is
possible within the formalism outlined below). The reservoirs are
homogeneous superconductors with a constant complex order parameter
except near the interface region. Within our model, the constriction
may be either superconducting or normal, although we will focus here
on the superconducting case.

We use a local orbital representation for
the description of the electronic states of the system. This
representation is very well suited for the self-consistent determination
of the order parameter profiles, provided that the pairing potential
is assumed to be diagonal in this basis. This simplification does
not affect in a relevant way the description of the superconducting state,
which is controlled basically by a single parameter (the coherence
length $\xi_0 \approx \hbar v_F / \pi \Delta$).
In the lower part of Fig. 1 we represent,
schematically, our discretized model constriction where the quasi one
dimensional region is represented by a linear chain with $N_C$ sites.
For convenience, in the numerical self-consistent calculations,
the left and right reservoirs are represented by two Bethe lattices with
coordination number $z+1$. The mean-field hamiltonian for this model has
the form

\begin{eqnarray}
H & = & \sum_{i,\sigma} (\epsilon_i - \mu)
c^{\dagger}_{i \sigma} c_{i \sigma}
+ \sum_{i \neq j, \sigma} t_{ij} c^{\dagger}_{i \sigma} c_{j \sigma}
\nonumber \\
& & + \sum_{i}( \Delta^{\ast}_{i} c^{\dagger}_{i \downarrow}
c^{\dagger}_{i \uparrow} + \Delta_{i} c_{i \uparrow} c_{i \downarrow}),
\end{eqnarray}
\noindent
where the chemical potential $\mu$ is a constant throughout the whole
system due to the absence of an applied voltage and will be taken as
zero. The hopping parameters
$t_{ij}$ are only different from zero for nearest neighbours and
the complex order parameters are given by

\begin{equation}
\Delta_{i} = -U_i <c^{\dagger}_{i \downarrow} c^{\dagger}_{i \uparrow}> ,
\end{equation}

\noindent
where $U_i$ is the attractive e-e interaction at site $i$. The system
defined by Eq. (2) provides the set of self-consistent conditions for
obtaining the order parameter profile.
 We choose $U_i$ and $t_{ij}$ to be constant
($U_{L}$, $U_{C}$, $U_{R}$ and $t_{L}$, $t_{C}$, $t_{R}$)
inside each different region (left reservoir, constriction, right
reservoir) in such a way as to fix the desired bulk values of $\Delta$
($\Delta^{0}_{L}, \Delta^{0}_{C}, \Delta^{0}_{R}$) in the three separate
regions. The hopping parameters, $t_{ij}$, can be chosen to be real and
we take $t_{L}=t_{R}=t$ as the unit of energy. We denote by $t_{LC}$ and
$t_{CR}$ the parameters coupling the left and right electrodes to the
constriction.

In the absence of an applied bias voltage, the Josephson current
through the constriction between sites $i$ and $i+1$ is given by

\begin{equation}
I_{i}(t) = \frac{i e}{\hbar} t_C \sum_{\sigma} \left( < c^{\dagger}_{i
\sigma}(t) c_{i+1 \sigma}(t) > - < c^{\dagger}_{i+1
\sigma}(t) c_{i \sigma}(t) > \right) .
\end{equation}
\noindent
One can show that $I_i$ is independent of the chosen site, i.e. the
continuity equation is fulfilled, only when the solution for the different
$\Delta_i$ is fully self-consistent \cite{Blonder,Sols}.

The averaged quantities appearing in Eqs. (2) and (3) are most conveniently
expressed in terms of non-equilibrium Green functions ${\bf G}^{+-} \;$
\cite{Keldish}, which in a Nambu (2x2)
representation \cite{Nambu} are defined by :

\begin{equation}
{\bf G}^{+-}_{i,j}(t,t^{\prime})= i \left(
\begin{array}{cc}
<c^{\dagger}_{j \uparrow}(t^{\prime}) c_{i \uparrow}(t)>   &
<c_{j \downarrow}(t^{\prime}) c_{i \uparrow}(t)>  \\
<c^{\dagger}_{j \uparrow}(t^{\prime}) c^{\dagger}_{i \downarrow}(t)>  &
<c_{j \downarrow}(t^{\prime}) c^{\dagger}_{i \downarrow}(t)>
\end{array}  \right) .
\end{equation}
\noindent
Then, the self-consistent Eqs. (2) and the stationary current, Eq.(3), are
given, in terms of the Fourier-transform ${\bf G}^{+-}_{i,j}(\omega)$, by

\begin{equation}
\Delta_i = -\frac{U_i}{2 \pi i} \int^{\infty}_{-\infty} d\omega [{\bf G}
^{+-}_{ii}(\omega)]_{21},
\end{equation}

\begin{equation}
I_i =\frac{2 e}{h} t_C \int^{\infty}_{-\infty} d\omega
\left( [{\bf G}^{+-}_{i+1,i}
(\omega)]_{11} -  [{\bf G}^{+-}_{i,i+1}(\omega)]_{11} \right).
\end{equation}
\noindent
All the different correlation functions appearing in Eqs.(5) and (6) can
be obtained using conventional Green function techniques \cite{GRGA}.

In the limit of $L_C \rightarrow 0$ it is possible to obtain analytical results
for the Josephson current using this model. In this case, the
self-consistent
phase profile is well approximated by a step function between the $L$ and $R$
electrodes. Defining $\phi = \phi_{L} - \phi_{R}$ as the total phase
diference, Eq.(6) can be written as

\begin{equation}
I = \frac{8 e}{h} t_{LR}^2 \sin(\phi) \int^{\infty}_{-\infty} d\omega
Im[ \frac{\tilde{g}^{r}_{L,21}(\omega) \tilde{g}^{r}_{R,12}(\omega)}{D^r
(\omega)} ] f(\omega).
\end{equation}

\noindent
where $t_{LR}$ is the coupling between the outermost
sites on the left and right reservoirs, $f(\omega)$ is the Fermi distribution
function;
${\bf g}^{r}_{L}$ and ${\bf g}^{r}_{R}$
are the retarded Green functions for the uncoupled electrodes (the tilde
indicates that the phase factor has been removed, i.e. $g^{r}_{L,21}= e^{i
\phi_{L}}\tilde{g}^{r}_{L,21}$ and $g^{r}_{R,12}=e^{-i \phi_{R}}
\tilde{g}^{r}_{R,12}$), and

\begin{equation}
D^{r}(\omega) = det[ {\bf I} - t_{LR}^2 {\bf \tau}_3 {\bf g}^{r}_{L}(\omega)
{\bf \tau}_3 {\bf g}^{r}_{R}(\omega) ],
\end{equation}

\noindent
${\bf \tau}_{3}$ being the usual Pauli matrix.

It is interesting to analyze the transition
from the tunnel to the contact
regime as given by Eq. (7) (for simplicity we assume
$\Delta^{0}_{L}=\Delta^{0}_{R}=\Delta$, and $\Delta \ll 1$ ). In the tunnel
regime,
$t_{LR} \ll 1$, $D^{r} \sim 1$,
and one recovers the usual Josephson expression for a tunnel junction
\cite{Ambe}

\begin{equation}
I(\phi) = \frac{\pi \Delta(T)}{2 e R_N} \sin \phi \tanh \left[
\frac{\Delta(T)}{2 k_B T} \right],
\end{equation}

\noindent
where $R_{N}$ is the normal resistance of the junction which
is given by $R^{-1}_{N}=(2e^{2}/h) \alpha$,
$\alpha$ being the normal
transmission through the constriction which in our model
adopts the simple form $ \alpha = (4
t^{2}_{LR}/ zt^{2})/(1+t_{LR}^{2}/zt^{2})^{2} $ \cite{contact}.
On the other hand, when approaching the
contact limit, $\alpha \rightarrow 1$,
the main contribution to the current comes from states inside the
superconducting gap, which are given by the zeros of Eq.(8).
These states are originated by multiple reflection processes at the
interface region and give the following simple expression for the current

\begin{equation}
I(\phi) = \frac{\pi}{2 e R_{N}} \frac{|\Delta(T)|^{2}}{|\epsilon(\phi)|}
\sin(\phi)
\tanh[\frac{|\epsilon(\phi)|}{2 k_B T}] ,
\end{equation}

\noindent
$\epsilon(\phi) = \pm |\Delta(T)| \sqrt{ 1 - \alpha \sin^2(\phi/2)}$
being the position of the
states inside the gap .
For the special case
$\alpha = 1$ Eq. (10) yields $I(\phi) \sim \sin(\phi/2)$, which coincides
with the result given in Refs. \cite{Beenakker,Jaime}.
It is remarkable that, when $\alpha \rightarrow 0$,
Eq.(10) tends exactly to Eq.(9), whose deduction involved no localized
states. One can understand the equivalence between
both ways of obtaining the tunnel limit  by realizing that  the
localized states move towards the gap edges
when $\alpha \rightarrow 0$, gradually becoming the
band edge singularities of the uncoupled system. These singularities
gave the main contribution to the current
when obtaining Eq.(9).

Let us turn our attention to the effects of self-consistency as one moves
from the above regime to the opposite case, $L_C /\xi_{0} \gg 1$. We have
performed calculations for a wide range of constriction lengths and
different values of the normal transmission parameter, $\alpha$.
The coordination number $z$ on both electrodes has been taken equal to
3, which ensures a rapid convergence to the bulk values of the order
parameter.
The self-consistent order parameter profile is calculated in the
following way: we first assume a given initial profile along the
constriction. Then, we calculate the system Green functions for this
profile which in turn yield the new order parameter profile using
Eq. (5). This process is repeated until convergence is achieved.

In order to clarify the $L_C / \xi_{0}$ dependence
of our results we have
fixed $\Delta^{0}_{C} = \Delta^{0}_{L} = \Delta^{0}_{R} = 0.05$ and
$\lambda_{F} = 4a$, ($a$ being the intersite distance), throughout the
whole calculation. For the same reason we restrict ourselves to the
zero temperature case in the present calculations.
In Fig.2 we show $I$ versus $\phi$ curves chosen
to represent typical behaviours found in the different physical regimes.
Thus, Fig.2a corresponds to the case of maximum transmission ($\alpha=1$)
, whereas Figs.2b and 2c illustrate cases with decreasing $\alpha$. In
these figures we plot curves for different values of
$L_C/ \xi_0$,
ranging from $L_C / \xi_{0}=0$ to $L_C / \xi_{0} \gg 1$. We only represent the
part of the curves with $I(\phi) \geq 0$.

Several features are noticeable in these curves. First, when
increasing $L_C / \xi_{0}$, a critical value
$L_C / \xi_0 \sim 1$ is reached
above which the function $I(\phi)$ becomes multivalued
\cite{Likharev,Ginzburg,Sols}. This
situation corresponds
to the appearance of a second kind, or ``solitonic",
solution with a phase slip center
inside the constriction, in addition to the normal one consisting
basically on a linear phase profile between the reservoirs.
In Figure 3 the
two kind of self-consistent profiles for $L_C > \xi_0$ are displayed.
They correspond to $\phi \simeq \pi$ where the phase drop inside
the constriction is specially sharp.
It is worth noticing the appearance of oscillations of period $\lambda_F /2$
in both profiles (phase and modulus).
This interference effect is a consequence of the phase coherence of the
normal
electrons along the constriction, analogous to the oscillations found in the
electrochemical
potential in normal mesoscopic wires \cite{Pernas,Buttiker-Pasta}.
On the other hand,
the other relevant length scale, $\xi_0$, manifests
itself in a clear way in the solitonic
phase profile: one can verify that the width of the phase-slip center is almost
equal
to $\xi_0$.
 The overall features of these solitonic solutions are in agreement with
predictions made in previous phenomenological analysis
\cite{Likharev,Baratoff}.

It is interesting to observe the behaviour of the maximum Josephson current,
$I_{max}$, as a function of $L_{C}$ for the different
situations represented in Fig. 2.
In the case of maximum transmission (Fig.2a) a slow
decrease in $I_{max}$ is observed for
increasing $L_C$, with a limiting value corresponding to
the critical
current of the infinite one-dimensional chain.
This monotonic decrease of $I_{max}$ is not found
when the trasmission is lowered below a certain value.
For instance, for the case of Fig.2b,
$I_{max}$ slightly increases with $L_C$, tending to the same limiting value as
in Fig.2a.

Fig. 2c corresponds to a somewhat different physical situation, in which a
weaker coupling between the constriction and the reservoirs is considered. For
sufficiently large $L_C$, this situation may be
regarded as a case not far from two
identical tunnel junctions connected in series, which would have a $ I(\phi)
\sim \sin (\phi / 2) $ characteristic, whereas
for $ L_C < \xi_0$ the
system behaves like a single junction with the usual $ I(\phi) \sim \sin \phi $
form. The transition between these two regimes can be clearly observed in fig.
2c. Note that with increasing $L_C$, $I(\phi)$ becomes eventually multivalued,
as in the previous cases. However, for large
enough $L_C$ (see case $N_C=80$ in fig. 2c), the
solitonic branch does not merge in a continuous way into the ``normal" one.
Instead, both branches extend up to
$\phi = 2 \pi$ keeping a different character. In the lower branch, the
region where the modulus of the order parameter nearly
goes to zero tends to fill the whole constriction when $\phi \rightarrow 2
\pi$.

In conclusion, we have presented a
fully self-consistent description of the d.c. Josephson
transport on a mesoscopic
weak-link. As a first check, the method has been applied to the case
of a short constriction, recovering previously known
results. On the other hand, we have analyzed the case where the
constriction length is comparable or larger than
the superconducting coherence length. To our knowledge,
this is the first calculation of the self-consistent
order parameter profiles and the current-phase relationship
based on a microscopic model.
The local character of our procedure makes it specially powerful for
treating systems with a more complicated geometry in situations where
the self-consistent variation of the order parameter plays a crucial role.
For instance, using the present approach one
would be able to address the problem of the proximity induced
Josephson effect in a
metal-superconductor nano-junction \cite{Agrait}. Work along this line
as well as
on the inclusion of a finite voltage for the study of the a.c.
Josephson effect is under progress.

Support by Spanish CICYT (contract no. PB89-0165) is acknowledged. The authors
are indebted to F. Flores, F. Sols, J. Ferrer and J.P. Hern\'andez for
stimulating discussions during the course of this work.

\begin{figure}
\caption{Schematic representation of the model constriction considered
in this paper.}
\end{figure}

\begin{figure}
\caption{Josephson current versus total phase difference for three different
values of
the transmission parameter $\alpha$: (a) $\alpha=1$, (b) $\alpha=3/4$, and (c)
$\alpha \simeq
1/12$. In cases (a) and (b) the hopping parameter between the constriction and
the reservoirs
($\protect t_{LC}$  and $\protect t_{CR}$) is equal to the intra-chain hopping
$t_C$
($t_C=\protect \sqrt{z} $
for case (a) and $t_C=1$ for case (b)), whereas in case (c) we take $t_C=1$
 and $\protect t_{LC}= \protect t_{CR}=0.5$. The numbers above each curve
indicate the number of sites within the constriction.}
\end{figure}

\begin{figure}
\caption{Self-consistent order parameter phase and modulus profiles for
the case $N_C=64$ and $\phi=3.4$ of Fig.2 (b). Solid and dotted
lines correspond to the solution in the upper and lower
branches of the $I(\phi)$ curve respectively.}
\end{figure}

\end{document}